\documentclass{article}
\usepackage{amsmath}
\usepackage{amsfonts}
\usepackage[latin1]{inputenc}
\newcommand{\di}{\displaystyle}

\newcommand{\R}{\mathbb  R}
\newcommand{\C}{\mathbb  C}
\newcommand{\N}{\mathbb  N}
\newcommand{\1}{\mathbb  I}

\newtheorem{lemma}{Lemma}[section]
\newtheorem{theorem}[lemma]{Theorem}
\newtheorem{proposition}[lemma]{Proposition}
\newtheorem{corollary}[lemma]{Corollary}
\newtheorem{remark}[lemma]{Remark}

\newtheorem{definition}[lemma]{Definition}

\def\sq{\hbox {\rlap{$\sqcap$}$\sqcup$}}
\def\sq{\hbox {\rlap{$\sqcap$}$\sqcup$}}
\def\noi{\noindent}

\def\beq{\begin{equation}}   \def\eeq{\end{equation}}
\def\bea{\begin{eqnarray}}  \def\eea{\end{eqnarray}}

\def\noi{\noindent}
\def\non{\nonumber}
\def\Tr{\rm Tr}

\newcommand\mysection{\setcounter{equation}{0}\section}
\renewcommand{\theequation}{\thesection.\arabic{equation}}
\newcounter{hran} \renewcommand{\thehran}{\thesection.\arabic{hran}}

\def\bmini{\setcounter{hran}{\value{equation}}
    \refstepcounter{hran}\setcounter{equation}{0}
    \renewcommand{\theequation}{\thehran\alph{equation}}\begin{eqnarray}}

\def\bminiG#1{\setcounter{hran}{\value{equation}}
\refstepcounter{hran}\setcounter{equation}{-1}
\renewcommand{\theequation}{\thehran\alph{equation}}
\refstepcounter{equation}\label{#1}\begin{eqnarray}}

\begin{document}

\title {Asymptotic expansion \\  for nonlinear eigenvalue problems}
\author{Fatima Aboud and Didier Robert\\
D\'epartement de Math\'ematiques\\
Laboratoire Jean Leray, CNRS-UMR 6629\\
 Universit\'e de Nantes, 2 rue de la Houssini\`ere, \\
F-44322 NANTES Cedex 03, France\\
\it didier.robert@univ-nantes.fr}
\vskip 1 truecm
\date{}
\maketitle

\begin{abstract}
In this paper we consider generalized eigenvalue problems for a family of operators
 with a quadratic  dependence on a complex parameter. Our model is 
 $L(\lambda)=-\triangle +(P(x)-\lambda)^2$ in $L^2(\R^d)$ where $P$ is a positive elliptic polynomial in $\R^d$
  of degree $m\geq 2$.
 It is known that for $d$ even, or $d=1$,  or $d=3$ and $m\geq 6$, there exist $\lambda\in\C$ and $u\in L^2(\R^d)$,
 $u\neq 0$, such that $L(\lambda)u=0$. In this paper,   we give a method to prove existence of non trivial solutions for the equation
  $L(\lambda)u=0$, valid in every dimension. This is a partial answer to a conjecture in \cite{herowa}.
\end{abstract}

\pagestyle{myheadings}

\mysection{Introduction}

             Let us introduce the following family of differential operators,
            \beq\label{eq5}
            L(\lambda) = -\triangle_x + \left(P(x) - \lambda\right)^2
            \eeq
            where $\triangle_x$ is the Laplace operator in $\R^d_x$,  $\lambda$ is a complex parameter,  $P$ is a  polynomial of degree $m\geq 2$ such that the leading homogeneous part $P_m$
             of $P$ satisfies $P_m(x) > 0$  for every $x\in \mathbb R^d\backslash\{0\}$
            (in other words we say that $P$ is a positive-elliptic polynomial). \\
            Such family of operators play an important role when studying analytic smoothness of solutions of  differential operators 
            with multiple characteristics (see \cite{herowa} and references there).  
            They also appear in the theory of damped oscillations in mechanics \cite{frsh, krla}. 
            The question we want to adress here is: ``does there exist
            $\lambda\in\C$
            and $u$ in the Schwartz space  ${\cal S}(\mathbb R^d)$, $u\neq 0$, such that $L(\lambda)u=0$?"
                      In \cite{chhela}  the authors have proven existence of non trivial solutions for 
            $1\leq d\leq 3$,  assuming that $m$  is large enough for $d=3$. After,
             Helffer-Robert-Wang  proved 
             in \cite{herowa}  the following result.
                \begin{theorem}
                Assume that $d$ is $\underline{even}$ and that $P$ is a positive-elliptic polynomial
                 of degree $m\geq 2$. \\
                 Then there exist  $\lambda\in \mathbb C$ and   $ u\in {\cal S}(\mathbb R^d)$, $u\neq 0$,
                  such that $L(\lambda)u = 0 $.
        \end{theorem}
       The proof given in \cite{herowa} shows that there exist  an infinite number of such eigenvalues \cite{ro4}
        located in the half-plane $\{\lambda\in \C,\;\; \Re\lambda\geq 0\}$.
        But it is not known if the generalized eigenfunctions span all the Hilbert space $L^2(\R^d)$, excepted  for $d=1$ \cite{phro}.\\
      For  $d$ odd , $d\geq 3$, $m\geq 2$,  the problem of existence of  non zero solutions is still open
      and it was conjecture in  \cite{herowa}  that such solutions exist whatever the dimension $d$. \\
      In this paper we  prove that this is true for every elliptic polynomial if $d=3$ and for large classes of elliptic polynomials  for $d=5, 7$.
      We  also discuss a numerical 
      approach to prove  that some coefficient in a semi-classical trace formula  is not zero. 
      For $d\geq 9$ we conjecture that this  coefficient
      is not zero hence  there exist an infinite number of nonlinear eigenvalues.
      
     {\em  This work was supported by the program ANR 08-BLAN-0228, NONAa,
     Research French Ministry.}
                      
       \section{nonlinear eigenvalue problems}
       In this section we recall some known  properties concerning nonlinear eigenvalue problems.
       For more details we refer to \cite{gokr, ma, ro4}.\\
       Let us consider the quadratic family of operators $L(\lambda)=L_0+\lambda L_1 +\lambda^2$
       where $L_0$, $L_1$ are operators in an Hilbert space ${\mathcal H}$.    
       $L_0$ is assumed to be self-adjoint, positive, with a domain $D(L_0)$
       and $L_1$ is $\sqrt{L_0}$-bounded. Moreover $L_0^{-1/2}$ is in a Schatten class
       ${\mathcal C}^p({\mathcal H})$ for some real $p>0$.   \\
          The following results are well known.
          \begin{theorem}
          $L(\lambda)$ is a family of closed operators in ${\mathcal H}$. \\
          $\lambda\mapsto L^{-1}(\lambda)$ is meromorphic in the complex plane.\\
          The poles  $\lambda_j$ of $L^{-1}(\lambda)$,   with multiplicity $m_j$, 
          co\"{\i}ncide with the eigenvalues with the same multiplicities, of the matrix
          operator  ${\mathcal A}_L$ in the Hilbert space ${\mathcal H}\times D(L_0^{1/2})$,
           with domain $D({\mathcal A}_L)=D(L_0)\times D(L_0^{1/2})$
           where 
             \beq
                           {\mathcal  A}_P = \left(
                            \begin{array}{cc}
                            0 & \1\\
                            -L_0&-L_1\\
                            \end{array}
                            \right).
                            \eeq
                            \end{theorem}   
                              Let us denote Sp[$L$] the eigenvalues of   ${\mathcal A}_L$ (which co\"{\i}ncide with the poles of $L^{-1}(z)$).
             \begin{remark}
             It may happens that Sp[$L$] is empty.  The following  one dimensional example is interesting and 
             was discussed in \cite{phro, chr1, chr2}.
             \beq
             L_{m,g}(\lambda) = -\frac{d^2}{dx^2} + (x^m-\lambda)^2 + gx^{m-1}.
             \eeq
             For every $m\geq 2$, $m$  even,   $L_{m,0}$ has infinity many eigenvalues but $L_{m,m}$ has no eigenvalue.
             The last statement is a consequence of the factorization
             $$
             L_{m,m}(\lambda) = (x^m-\lambda +\frac{d}{dx})(x^m-\lambda -\frac{d}{dx}).
             $$
             So, we can compute all solutions for the equation $L_{m,m}(\lambda)u=0$ and see that a non-null
             solution   $u$ is never bounded on $\R$. \\
             But if $m$ is odd,  $L_{m,m}(\lambda)u=0$,  has infinity many eigenvalues on the imaginary axis
             \cite{chr2}.\\
             On the other side there exist  sufficient  general conditions  to have  
             ${\rm Sp}[L]\neq \emptyset$ \cite{gokr, ma}. Unfortunately these conditions are not fulfilled for our
             example $L(\lambda) =  -\triangle_x +(P(x)-\lambda)^2$  when $d\geq 2$.
             \end{remark}
        The following formula appears  for the first time in \cite{boko} and will be very useful for our purpose.
                    \begin{theorem} For $k$ large enough ($k\in\N$, $k>p$) and for $z\in\C\backslash{\rm Sp}[L]$,
                            we have
                            \beq\label{bk}
                            {\rm Tr}({\mathcal A}_L-z)^{-k-1} =
                             \frac{-1}{k!}{\rm Tr}[\frac{d^k}{dz^k}(L(z)^{-1}L^\prime(z)],
                            \eeq
                            where each above operators are  trace class.
                            \end{theorem}
                            Using Lidskii Theorem \cite{gokr} and (\ref{bk}),  we get 
                              \beq\label{Lid1}
                            \sum_{\lambda\in{\rm Sp}[L]}m_\lambda(\lambda-z)^{-k-1} = 
                            \frac{-1}{k!}{\rm Tr}[\frac{d^k}{dz^k}\left(L(z)^{-1}L^\prime(z))\right)].
                            \eeq
                            where $m(\lambda)$   is the multiplicity of the eigenvalue $\lambda$.

                           As it was nicely remarked  in the paper \cite{chhela}, a sufficient condition for ${\rm Sp}[L]\neq \emptyset$ 
                           is that the r.h.s in (\ref{Lid1}) is not zero.  To check this property  a natural method is to introduce parameters
                           and use semiclassical analysis.\\
                           In \cite{herowa} the authors also use Lidskii theorem and  semi-classical analysis on the matrix system 
                           ${\mathcal A}_L$.
                          Here we consider  more directly  the scalar family of operators $L(z)$ where computations are easier even if the dependence
                          in $z$ is nonlinear.
     
     \section{ Semiclassical parametrix}
   For simplicity
      we assume  here that $P$ is homogeneous of degree $m\geq 2$ and $P(x)>0$ for $x\in\R^d$, $x\neq 0$.   
      By the  scaling transformation 
      $x = \tau^{1/m}y$ with $\hbar = \tau^{-(m+1)/m}$ and 
      $z = \dfrac{\lambda}{\tau} $ we  can  see that $L(\lambda)$
       is unitary equivalent to the semiclassical Hamiltonian
        $\tau^2\hat L(z)$ where 
        \beq\label{eq9}
        \hat L(z) = -\hbar^2\triangle_x + \left(P(x)  -z\right)^2.
        \eeq
         $ \hat L(z) $ is the $\hbar$-Weyl operator with the symbol
         $L(z, x, \xi) = \xi^2 +  \left(P(x)  -z\right)^2$. 
        For semiclassical analysis tools and  $\hbar$-Weyl quantization we refer to \cite{ro2}.
                 Here we use the notation $\hat H$ for the  $\hbar$-Weyl quantization
         of the symbol $H$ or  for convenience, $\hat H = Op^w_\hbar(H)$.
        
        Using semiclassical operator calculus, we  can  construct a good parametrix for $\hat L(z)^{-1}$
        for $z\in\Lambda$ where $\Lambda$ is the sector
        $$
        \Lambda = \{z\in\C,\; \vert z\vert \geq r_0,\;  \pi/2+\delta<\arg(z)<3\pi/2-\delta\};\;\; r_0>0, \; \delta>0.
        $$
           \begin{theorem} There exists a semiclassical symbol $K^{(\hbar)}(z)$, $z\in\Lambda$, $0<\hbar<1$, such that
           \bea\label{param}
           K^{(\hbar)}(z;x,\xi) &\asymp& \sum_{j\geq 0} \hbar^{2j}K_{2j}(z;x,\xi), \nonumber\\
           \hat L(z)^{-1} &=& Op^w_\hbar(K_\hbar(z)).
           \eea
           Moreover the asymptotic expansion has the following meaning: 
           for every $N\geq 1$  we have 
           $$
           \hat L(z).Op^w_\hbar\left(\sum_{0\leq j\leq N} \hbar^{2j}K_{2j}(z)\right)  =
            \1 + \hbar^{2N+2}Op^w_\hbar\bigl(R^{(\hbar)}_{2N}(z)\bigr)
           $$
           where the symbol $R^{(\hbar)}_{2N}(z)$ satisfies the following estimates :\\
           for every 
           $\alpha, \beta\in\N^d$ we have
           \beq\label{paramest}
           \Bigl\vert\partial^\alpha_x\partial^\beta_\xi\left(R^{(\hbar)}_{2N}(z;x,\xi)\right)\Bigr\vert \leq 
       C(N,\alpha,\beta)\frac{\mu(x,\xi)^{2m}+\vert z\vert\mu(x,\xi)^{m}}{\mu(x,\xi)^{2m} +\vert z\vert^2}
       \mu(x,\xi)^{-2N-\vert\alpha\vert-\vert\beta\vert}
           \eeq
           for every 
           $\alpha, \beta\in\N^d$, where  $C(N,\alpha,\beta)$ is uniform in $z\in\Lambda$ and 
           where $\mu(x,\xi)=(1+\vert x\vert^{2m}+\vert\xi\vert^2)^{1/2m}$.
           \end{theorem}    
              {\bf sketch of proof.}   The method to get such result is standard and  was used many times to construct parametrix 
              of elliptic pseudo-differential operators \cite{se}.      Usually the $z$-dependence  is linear but here it is quadratic.  Moreover here we need accurate estimates for the remainder term in the product of
              pseudo-differential operators depending on parameters. The necessary estimates for 
                $R^{(\hbar)}_{2N}(z;x,\xi)$  are
              established using   the technics coming  from the papers  \cite{daro, boro}. \\
              An other difficulty here  is that  we shall need to compute the symbols $K_{2j}$ for $j$ large enough. 
                         This computations are not easy, so we have to be explicite as far as possible.\\
              Using the  product formula for $\hbar$-pseudodifferential operators, we get at the initial step: 
              \beq\label{stp0}
              K_0(z;x,\xi) = \frac{1}{L(z;x,\xi)} = \frac{1}{\vert\xi\vert^2+(P(x)-z)^2}
              \eeq
        and the induction formula
        \beq\label{stp2j}
        K_{2j} = -K_0\left(\sum_{0\leq\ell\leq j-1}\sum_{\vert\alpha\vert+\vert\beta\vert=2(j-\ell)}\Gamma(\alpha,\beta)
        \partial_\xi^\alpha\partial_x^\beta L(z)\partial_\xi^\beta\partial_x^\alpha K_{2\ell}\right)
        \eeq
        where $\Gamma(\alpha,\beta)= \frac{(-1)^{\vert\beta\vert}}{2^{2(j-\ell)}\alpha!\beta!}$.
        Let us compute $K_2$ and $K_4$.\\
        \bea\label{stp2}
        K_2 &=& \frac{L_2(z)}{L^3(z)} + \frac{L_3(z)}{L^4(z)}, \nonumber\\
        L_2(z) &=& (P(x)-z)\triangle P(x) +\vert\nabla P(x)\vert^2, \nonumber \\
        L_3(z) &=& -2[(P(x)-z)D^2P(x)\xi\cdot\xi + (\nabla P(x)\cdot\xi)^2 + (P(x)-z)^2\vert\nabla P(x)\vert^2 ],\nonumber
        \eea 
        where $D^2P(x)$ is the Hessian matrix of $P$ in variable $x$.\\
       Now using (\ref{stp2j})  we have 
       \bea\label{stp4}
K_4 = & -K_0\Bigl\lbrace  \sum_{ \vert\beta\vert=4 } \Gamma(0,\beta)\partial_x^\beta L(z)\partial_\xi^\beta K_0 
        + \sum_{\vert\alpha\vert=2}\Gamma(\alpha, 0)\partial_\xi^\alpha L(z)\partial_x^\alpha K_2 \nonumber\\
        & +  \sum_{\vert\beta\vert=2}\Gamma(0,\beta)\partial_x^\beta L(z)\partial_\xi^\beta K_2)  \Bigr\rbrace.
       \eea
      By induction on $j$, we  easily get that 
      \beq\label{step2jp}
      K_{2j}(z;x,\xi) = \sum_{j+1\leq k\leq 3j}\frac{Q_{k}^{2j}(x,P-z,\xi)}{L(z;x,\xi)^{k+1}},
      \eeq
        $Q_{k}^{2j}(x,P-z,\xi)$ is  a  polynomial in $((P-z),\xi)$, with a total degree $\leq k-2$, 
         with coefficients  depending on derivatives of $P(x)$. \\
         The following lemma will be useful later.
         Let us denote val[$Q_{k}^{2j}$], the valuation of  $Q_{k}^{2j}$ as a polynomial in $P-z, \xi$. 
         Let us recall the definition of valuation.
          Denote by ${\cal I}$ the ideal with generators $\xi_1,\cdots, \xi_d, P-z$,   in the ring $C^{\infty}(\R_\xi\times\R_x)$.
         If $Q\in C^{\infty}(\R_\xi\times\R_x)$, val$[Q] $ is the biggest integer $p$ such that $Q\in{\cal I}^p$. 
         \begin{lemma}\label{val}
         We have
         $$
         {\rm val}[Q_{k}^{2j}] \geq 2(k-1-2j),\;\; {\rm for}\;\; 2j+2 \leq k\leq 3j,\; {\rm and}\;\; j\geq 1.
         $$
         \end{lemma}
         {\bf Proof}. 
      This is easily proved by induction on $j$, using  (\ref{step2jp})  and the following formula.
      Let $Q$ and $L$  be smooth functions in $\R^n$, a multiindex $\alpha\in\N^n$, then we have
      \beq
      \partial^\alpha\left(\frac{Q}{L^{k+1}}\right) =  \sum C(\mu_j,\gamma_k)
      \frac{\partial^{\alpha-\gamma}Q(\partial^{\gamma_1}L)^{\mu_1}\cdots (\partial^{\gamma_\ell}L)^{\mu_\ell}}{L^{\mu+k+1}}
       \eeq
      where in the sum we have the conditions, $\gamma_j\in\N^n$, $\mu_j\in\N$,  
      $\gamma \leq \alpha$, $\mu_1+\cdots\mu_\ell=\mu$,
      $\mu_1\vert\gamma_1\vert+\cdots+\mu_\ell\vert\gamma_\ell\vert=\vert\gamma\vert$.
       \sq
       \begin{remark}
       The parametrix computed above is enough to get qualitative informations. 
       Quantitative informations are much more difficult to get
       except for the first  orders ($j =0,1$). When $j$ is larger it is not so easy to compute explicitely the terms
         $Q_{k}^{2j}(x,P-z,\xi)$.
       \end{remark}
       \begin{remark}\label{polyh}
       It is not difficult to extend the above results when the elliptic polynomial $P(x)$ has lower terms:
       $P = P_m+P_{m-1}+\cdots P_1+P_0$ where $P_j$ is homogeous with degree $j$ and
       $P_m(x)>0$ for $x\in\R^d\backslash\{0\}$.  Then we have
       $$
       P(\tau^{1/m}y)=\tau P^{(\varepsilon)}(y)
       $$
       with $\varepsilon =  \tau^{-1/m} = \hbar^{1/(m+1)}$ and 
       $P^{(\varepsilon)}(y) = P_m(y) +\varepsilon P_{m-1}(y)+\cdots+\varepsilon^mP_0(y)$. So $P^{(\varepsilon)}$ is a uniform elliptic family
       of  polynomials and we can easily see that the constructions  in (\ref{paramest})    are uniform in the small parameter $\varepsilon$.   
       \end{remark}
        \section{A trace formula}
        Recall that ${\rm Sp}[L]$ denote the generalized eigenvalues of the quadratic family $L(z)$, $m_\lambda$ is the multiplicity of the eigenvalue $\lambda$. Let $f$ an holomorphic function in $\Lambda$ such that 
      \beq\label{hypf}
        \vert f(z)\vert \leq C(1+\vert z\vert)^{-\mu},\;\;   \forall z\in\Lambda.
       \eeq
       For our applications we shall choose $f(z) = (z+\lambda)^{-\mu}$, for a suitable parameter $\lambda\in\C$.
       Let be $\Gamma$ a complex contour in $\Lambda$ defined as follows.
       $$
       \Gamma = \{ r{\rm e}^{\pm i\theta_0},\; r\geq r_0\}\cup\{ r_0{\rm e}^{i\theta},\; \theta_0\leq \theta\leq 2\pi-\theta_0\},
       $$
       where $r_0>0$ and $\frac{\pi}{2}<\theta_0<\pi$.
   \begin{proposition}
   Assume that $\mu> \frac{d(m+1)}{m}$. Then  $f({\mathcal A}_L)$ is a trace class operator and we have
   \beq
   \Tr(f({\mathcal A}_L) = \sum_{\lambda\in{\rm Sp}[L]}m_\lambda f(\lambda) =
   \Tr[\oint_\Gamma \hat L(z)^{-1}\hat L^\prime(z)f(z)dz],
   \eeq
   where $\oint_\Gamma F(z)dz = \frac{1}{2i\pi}\int_\Gamma F(z)dz$ (contour integral in the  complex plane).
  \end{proposition}
{\bf Proof.} This a  direct  consequence of the Cauchy integral formula and Theorem \ref{bk}. \sq
   \begin{theorem}
   For $f$ as above, for every $d\geq 1$ we have in the  semiclassic regime $\hbar\searrow 0$,  modulo 
   ${\cal O}(\hbar^{+\infty})$, 
   \beq\label{trasymp}
    \sum_{\lambda\in{\rm Sp}[L]}m_\lambda f(\lambda) \asymp \sum_{j\geq 0}C_{2j}^{(d)}(f)\hbar^{2j-d}.
   \eeq
If $d$ is odd,  
   \beq\label{t00}
   C^{(d)}_0(f) =  0
   \eeq
   and for $d$ even, 
   \beq\label{t01}
   C^{(d)}_0(f) = 2(-1)^{d/2}(2\pi)^{-d}\int\!\!\int_{\R^{2d}}f(P(x)+\vert\eta\vert)dxd\eta.
   \eeq   
   For the other terms ($j\geq 1$) we have the following qualitative information
   \beq\label{t2j}
   C^{(d)}_{2j}(f) = \sum_{0\leq k\leq n_j}\int_{\R^d}A_{2j,k}(x)f^{(k)}(P(x))dx,
   \eeq 
   where  $A_{2j,k}(x)$ are polynomials in $\partial_x^\gamma P(x)$, $\vert\gamma\vert\leq 2j$
    and $n_j$ depends on $j$.\\
  \fbox{ Moreover if $d$ is odd, then  $C^{(d)}_{2j}(f)=0$ for $d\geq 4j+1$}.
   \end{theorem}   
      {\bf Proof.}
    The asymptotic expansion (\ref{trasymp})  is a  direct consequence  of (\ref{paramest}) and usual properties of trace 
     operation for Weyl quantization.\\
    Let us compute $   C^{(d)}_0(f)$. We have the integral formula:
    $$
       C^{(d)}_0(f) = -\oint_\Gamma\frac{2(P(x)-z)}{\vert\xi\vert^2+(P(x)-z)^2}f(z)dzd\xi\tilde dx,
       $$
       where $\tilde dx=(2\pi)^{-d}dx$.  By the residue theorem we get
       $$
       C^{(d)}_0(f) = \int\!\!\int_{\R^d\times\R^d}[f(P(x)+i\vert\xi\vert) + f(P(x)-i\vert\xi\vert)]d\xi\tilde dx.
       $$
       For $a>0$ we have
       $$
       \int\!\!\int_{\R^d\times\R^d}(f(P(x)+a\vert\xi\vert)d\xi\tilde dx= a^{-d}\int\!\!\int_{\R^d\times\R^d}(f(P(x)+\vert\xi\vert)d\xi\tilde dx.
$$
So by analytic extension and evalution at $a=i$ we get formula (\ref{t00}) and  (\ref{t01}). In particular we  see that for
$d$ even, there exists $f$ satisfying (\ref{hypf}) such that $C^{(d)}_0(f)\neq 0$.

For $ j\geq 1$, using (\ref{step2jp}),  we have 
\beq\label{C2j}
C^{(d)}_{2j}(f) = \int\!\!\int\oint_\Gamma\sum_{j+1\leq k\leq 3j}\frac{2(P(x)-z)Q_{k}^{2j}(x,P(x)-z,\xi)}{L(z;x,\xi)^{k+1}}
f(z)dzd\xi\tilde dx.
\eeq
Let us  now  prove that $C^{(d)}_{2j}(f) =0$, for $4j+1\leq d$.\\
To do that it is convenient to introduce the following integral, for $u>0, v>0$, 
\beq
J_{k,\nu}f(u,v) = \oint_\Gamma\frac{(v-z)^\nu}{(u+(v-z)^2)^{k+1}}f(z)dz.
\eeq
We have easily
\beq\label{J1}
J_{k,\nu}f(u,v) =\frac{(-1)^{k}}{k!}\frac{\partial^k}{\partial u^k}J_{0,\nu}f(u,v).
 \eeq   
   And using the residue theorem, we get
   \beq\label{J2}
    J_{0,\nu}f(u,v) = \frac{ i^{\nu-1}u^{(\nu-1)/2}}{2}\Bigl((-1)^{\nu+1}f(v+i\sqrt{u}) + f(v-i\sqrt{u})\Bigr).
    \eeq
 From (\ref{J1}) and (\ref{J2}) we can compute    $J_{k,\nu}f(u,v)$.\\
 To prove that $C^{(d)}_{2j}(f)=0$  for $d\geq 4j+1$, we shall prove that each term in the sum (\ref{C2j}) vanishes,
 after integration  in $z$ and $\xi$.\\
 Suppose first that $j+1\leq k\leq 2j+1$. We have
 $$
 Q_{k}^{2j}(x,P(x)-z,\xi) = \sum_{\nu, \gamma}R_{\nu,\gamma}(x)(P(x)-z)^{\nu}\xi^{\gamma}.
 $$
 Hence
 $$
 \int\oint_\Gamma\frac{2(P(x)-z)Q_{k}^{2j}(x,P(x)-z,\xi)}{L(z;x,\xi)^{k+1}}f(z)dzd\xi
 $$
 is a sum of integrals like
 $$
 I^k_{\nu}(f)(x,\xi) = \oint_\Gamma\frac{(P(x)-z)^{\nu}}{L(z;x,\xi)^{k+1}}f(z)dz.
 $$
    By integration by parts in $z$ we have
    \beq
     I^k_{\nu+1}(f) =  \frac{\nu}{2k}I^{k-1}_{\nu-1}(f) -  \frac{1}{2k}I^{k-1}_{\nu}(f^\prime).
     \eeq
     So, we  can assume that $\nu=0$. But we have
     $$
     I^\ell_0(g)(x,\xi) = \frac{(-1)^{\ell}}{\ell!}\frac{\partial ^\ell}{\partial u^\ell}J_{0,\nu}g(u,P(x))\vert_{u=\vert\xi\vert^2}.
     $$
    So  we have 
     $ I^\ell_0(g)(x,\xi) = {\cal O}(\vert\xi\vert^{2-2\ell})$ near $\xi=0$. 
      Now we remark that for $\ell\leq 2j+1$  and $d\geq 4j+1$ we have $\ell<\frac{d}{2}+1$, hence $\xi\mapsto I^\ell_0(g)(x,\xi)$
      is integrable and,  using the analytic dilation argument already used for $j=0$, we get 
      $  I^\ell_0(g)(x,\xi)=0$, hence  
  $$
 \int\!\!\oint_\Gamma\frac{2(P(x)-z)Q_{k}^{2j}(x,P(x)-z,\xi)}{L(z;x,\xi)^{k+1}}f(z)dzd\xi = 0.
 $$
    Now, assume that $2j+2\leq k\leq 3j$. Using Lemma.\ref{val},  we have
    $$
 Q_{k}^{2j}(x,P(x)-z,\xi) = \sum_{\nu+\vert\gamma\vert\geq2(2k-1-2j)}R_{\nu,\gamma}(x)(P(x)-z)^{\nu}\xi^{\gamma}.
 $$
As above, we integrate by parts in $z$ to have the possibility to put $\nu$ at 0 and then we use $\xi^\gamma$ to decrease the order
of the singularity in $\xi$ as far as possible (integrability near $\xi=0$) of $\oint_\Gamma\frac{Q_{k}^{2j}}{L^{k+1}}dz$.
We conclude by the analytic dilation argument. \sq 
    
    So, we have proven  that in odd dimension $d$,  $C^{(d)}_{2j}(f) = 0$ if $2j\leq \frac{d-1}{2}$.\\
   We conjecture that the next following  terms are  not 0; more precisely we claim:\\
   {\bf Conjecture:} For every $j\in\N$, $j\geq 1$, there exists $f$ satisfying (\ref{hypf}) such that we have  we have
   \beq
   C_{2j}^{(4j-1)}(f)\neq 0,\;\;{\rm and}\;\; C_{2j}^{(4j-3)}(f)\neq 0
   \eeq
    In the following sections we shall check this conjecture for 
    $d=1, 3$ and we shall compute analytic formula for $C_4^{(5)}(f)$ and $C_4^{(7)}(f)$. 
    Unfortunately, these analytic expressions have many terms and it is not obvious that $C_4^{(d)}(f)\neq 0$   for $d=5, 7$,
    for every elliptic polynomial $P$. We shall see that this is true for convex polynomials for $d=7$ and  satisfying a technical condition if $d=5$.
    Moreover we get,   using numerical computations  for particular non-convex polynomials $P$, 
    that $C_4^{(d)}(f)\neq 0$.\\
     As  we shall see in the next section, the property $ C_{2j}^{(d)}(f)\neq 0$  gives easily a lower bounds on the density of eigenvalues.
      \begin{remark}
     Following Remark \ref{polyh} we can extend our results to polyhomogeneous 
     polynomials $P  = P_m+P_{m-1}+\cdots P_1+P_0$. To follow the dependence
     in  the coefficients, we note $C_{2j}(f,P)$ the coefficient $C_{2j}(f)$ with polynomial
      $P$. \\
     In particular we have
     $C^{(d)}_{2j}(f, P^{(\varepsilon)}) = 0$, for $d\geq 4j+1$ and  for every $\varepsilon$ small enough.  Assume now that $d=4j_0-3$ or $d =4j_0-1$, $j_0\geq 1$. 
     Then using a Taylor expansion in $\varepsilon$,   computed for $\varepsilon = \hbar^{1/(m+1)}$,
     we get 
     \beq
     C_{2j}(f, P^{(\hbar^{1/(m+1)})}) \asymp
       \sum_{k\geq 0}\gamma_k\hbar^{k/(m+1)},
      \eeq
      in particular $\gamma_0 = C_{2j_0}^{(d)}(f,P_m)$ which is supposed to be not 0,
      as we have explained  before.

     \end{remark}
     \section{Estimate  the density of eigenvalues}
     First of all let us remark that the nonlinear spectrum ${\rm Sp}[\hat L]$ of $\hat L$ is included
     in the two  quarters $\{z\in\C,\; \Re(z)\geq 0,\;\; \pm\Im(z)>0\}$. \\
     On one side, it  is easy to see that if  $\lambda\in\R$ and $L(\lambda)u=0$ then $u=0$.  
     On the other side, if $\Re(\lambda)<0$ and $L(\lambda)u=0$, computing
     $\Im(\langle L(\lambda)u,u\rangle)$ we conclude that $u=0$.

    Let us denote by $N_\hbar(R) = \#\{z\in {\rm Sp}[\hat L];\; \vert z\vert \leq R\}$ and $N(R) = N_{\hbar =1}(R)$.
    \begin{proposition}
   For every real  $\mu$,  $\mu>d(m+1)/m$,  there exists $C_k>0$ such that 
   \beq\label{ub}
    N_\hbar(R) \leq C_kR^\mu\hbar^{-d},\;\; \forall R\geq 1, \;\forall \hbar\in]0, 1].
    \eeq 
    If  $ C_{2j}^{(d)}(f)\neq 0$ with $d>2j$,  then for every $r>0$, $\varepsilon>0$ there exists  $c_{\varepsilon, r}>0$ such that
   \beq\label{lb}
    N_\hbar(r\hbar^{-\varepsilon}) \geq c_{\varepsilon, r}\hbar^{-\delta},\;\; \forall \hbar\in]0, 1],
    \eeq
    where $\delta = d-2j$.
    Moreover if $j=0$ ($d$ even) then the estimates is valid with $\varepsilon =0$.
    So that, in even dimension,  for every $R>0$, $N_\hbar(R)$ 
     behaves like $\hbar^{-d}$. 
    \end{proposition}
    {\bf Proof.}
    The proof of (\ref{ub}) is a  direct consequence of  Weyl-Ky-Fan inequality  \cite{ro4}.\\
    We first remark that for every $\varepsilon >0$ there exists $R_\varepsilon>0$ such that if
   $$
    -\pi/2 -\varepsilon \leq \arg z \leq \pi/2 + \varepsilon, \;\:\; \vert\mu\vert \geq R_\varepsilon
    $$
    then we have
    $$
    \vert t+  z\vert^2 \geq (1-\varepsilon)(t^2+\vert z\vert^2).
    $$
  Let us  choose $f(\lambda) = (\lambda + t)^{-\mu}$ with $k$ large enough ($\mu>d(m+1)/m$) and $t>0$.
 We apply (\ref{trasymp})  to get the following inequalities
    $$
    C_1\hbar^{-\delta} \leq \bigl\vert\sum_{z\in Sp[\hat L]}(t+ z)^{-\mu}\bigr\vert 
    \leq \sum_{z\in Sp[\hat L]}\vert t+ z\vert^{-\mu} 
    \leq C_2 \sum_{z\in Sp[\hat L]}(t+\vert z\vert)^{-\mu}
    $$
    But for every $\mu, \mu_1$, large enough, such that $\mu-\mu_1$ is large enough, we have 
    $$
    \sum_{\frac{z\in Sp[\hat L]}{\vert z\vert\geq R}}(t+\vert z\vert)^{-\mu} 
    \leq R^{-\mu_1}\sum_{z\in Sp[\hat L]}(t+\vert z\vert)^{\mu_1-\mu}
    $$
    We choose now $R=r\hbar^{-\varepsilon}$ to get
    $$
     N_\hbar(r\hbar^{-\varepsilon}) \geq \sum_{\vert\mu\vert\leq R}(1+\vert\mu\vert)^{-k}\geq c_{\varepsilon, r}\hbar^{-\delta}
     $$
     \sq

     The above results concern the semi-classical regime. Now we give estimates for $\hbar =1$ and high energy regime
     \begin{corollary} For $R\nearrow +\infty$ we have
     $$
     N(R) = {\cal O}(R^{d(m+1)/m}).
     $$
     If    $ C_{2j}^{(d)}(f)\neq 0$ with $d-2j>0$, then for every $\varepsilon >0$ there exits $c_\varepsilon>0$ such that
     $$
     c_\varepsilon R^{\delta(m+1)/m-\varepsilon} \leq  N(R)
     $$
     If $j=0$, the estimate is true with $\varepsilon=0$ and $c_0>0$.
     \end{corollary}
      \section{ 1-$d$  and 3-$d$  cases}
      In this section we prove the following result.
      \begin{theorem}
      For  $d=1, 3$, there exits $f$ satisfying (\ref{hypf}) such that  for every $m\geq 2$, we have
            $C^{(d)}_{2}(f) \neq 0$.
      More precisely, we have
      \bea
      C^{(1)}_{2}(f) &=& -\frac{1}{16}\int_\R f^{(3)}(P(x))P^\prime(x)^2dx\\
      C^{(3)}_{2}(f) &=& -\frac{1}{48\pi}\int_{\R^3}f^\prime(P(x))\vert\nabla P(x)\vert^2dx
      \eea
      We can choose $f(\lambda) = (\lambda + t)^{-\mu}$ with $\mu>d(m+1)/m$ and $t>0$.
      \end{theorem}
      {\bf Proof.}  We compute with the explicite form we got before for $K_2$. We have,
      for $d=1, 3$, 
      $$
      C^{d}_2(f,x) = -2\oint_\Gamma (P-z)H_2dz,
      $$
      where 
      $\di{H_2 =\int_{\R^d}K_2d\xi}$. But we have
      $$
      H_2 = (P-z)^{d-5}\triangle P\Bigl(b_3 -2b_{4,1}\Bigr) + 
      (P-z)^{d-6}\vert\nabla P\vert^2\Bigl(b_3 -2b_4 - 2b_{4,1}\Bigr),
      $$
    hence
    \beq
      C_2^{(1)}(f) = \frac{1}{2\pi}\Bigl(\frac{8}{3}b_{4,1} -
      \frac{4}{3}b_{3,0} + \frac{2}{3}b_4\Bigr)\int_{\R}f^{(3)}(P(x))P^\prime(x)^2dx.
      \eeq
      \beq
     \boxed{ C_2^{(1)}(f) = -\frac{1}{16}\int_{\R}f^{(3)}(P(x))P^\prime(x)^2dx}
      \eeq
   \beq
       C_2^{(3)}(f) = \frac{2}{(2\pi)^3}\Bigl(4b_{4,1} -2b_4 -2b_3\Bigr)\int_{R^3}f^\prime(P(x))\vert\nabla P\vert^2(x)dx
      \eeq
and
\beq
    \boxed{C_2^{(3)}(f) = -\frac{1}{48\pi}\int_{\R^3}f^\prime(P(x))\vert\nabla P\vert^2(x)dx}
      \eeq

            \sq

        We have seen that for $d$ odd,  $d\geq 5$,  $C^{(d)}_{2}(f) = 0$.
         So we have to compute $C^{(d)}_{4}(f)$
         for $d=5,7$.
        
        \section{5-$d$ and 7-$d$ cases}
        We have to compute  in more details the term $K_4$ from (\ref{stp4}).  Recall that we have
        \beq
        C_4^{(d)}(f) = -2\int_{\R_x^d}\!\!\oint_\Gamma(P-z)\left(\int_{\R_\xi^d}K_4(z;x,\xi)d\xi\right)f(z)dz\tilde dx.
        \eeq
      We have to compute the following three integrals, depending on $x\in\R^d$ and $z\in \C$.
      \bea
      I_\beta^{(1)} &=& \Gamma(0;\beta)\int_{\R^d}\frac{1}{\vert\xi\vert^2+(P(x)-z)^2}
      \partial^\beta_\xi\left( \frac{1}{\vert\xi\vert^2+(P(x)-z)^2} \right)d\xi;\nonumber
   \vert\beta\vert=4,\\
       I_\alpha^{(2)} &=& \Gamma(\alpha,0)\int_{\R^d}\frac{1}{\vert\xi\vert^2+(P(x)-z)^2}\partial^\alpha_\xi(\vert\xi\vert^2)\partial_x^\alpha K_2d\xi;\;\;\; \;\; \vert\alpha\vert =2,\\
 I_\beta^{(3)} &=& \Gamma(0;\beta)\int_{\R^d}\frac{1}{\vert\xi\vert^2+(P(x)-z)^2}\partial^\beta_\xi K_2 d\xi;\;
      \;\; \vert\beta\vert=2.\\
       \eea
      Using the new variable $\eta$ such that $\xi=(P-z)\eta$ (plus an analytic extension), we get
      \beq
      I_\beta^{(1)}  = \frac{a(\beta)}{(P(x)-z)^{8-d}},\;\; a(\beta) = 
     \frac{1}{16\beta!} \int_{\R^d}\frac{1}{1+\vert\eta\vert^2}\partial_\eta^\beta\left(\frac{1}{1+\vert\eta\vert^2}\right)d\eta.
      \eeq
      $a(\beta)\neq 0$ only when $\beta=(\beta_1,\cdots, \beta_d)$ is such that $\beta_j=4$ and $\beta_k=0$
      for $k\neq j$ or $\beta_j=\beta_k=2$, $j\neq k$ and $\beta_\ell =0$ if $\ell\neq j$, $\ell\neq k$.\\
      In the first case $a(\beta)=a_1$ and in the second case $a(\beta)=a_2$ where
      \bea
      a_1 &=& \frac{1}{96}\int_{\R^d}\frac{(4\eta_1^2-(1+\vert\eta\vert^2))^2}{(1+\vert\eta\vert^2)^6}d\eta,\\
      a_2 &=&\int_{\R^d}\frac{\eta_1^2\eta_2^2}{(1+\vert\eta\vert^2)^6}d\eta.
      \eea
      It is convenient to introduce the following  notations.
      \bea
      b_j = \int_{\R^d}\frac{d\eta}{(1+\vert\eta\vert^2)^j},\;\;
       b_{j,k} =  \int_{\R^d}\frac{\eta_1^{2k}d\eta}{(1+\vert\eta\vert^2)^j}, \;\;
      b_{j,k,\ell} =  \int_{\R^d}\frac{\eta_1^{2k}\eta_2^{2\ell}d\eta}{(1+\vert\eta\vert^2)^j}. 
      \eea
      where $j,k,\ell \in\N$ are such that the integrals are finite. 
      Of course these integrals  can be computed with the Euler beta and gamma special functions (see appendix for more explicite expressions). \\
      So we have 
      $a_1 = \frac{1}{6}b_{6,2} -\frac{1}{3}b_{5,1}+\frac{1}{96}b_4$ and $a_2=b_{6,1,1}$.\\
     Using integration by parts, in $x$ or in $\xi$, we get the following formulas
      \beq\label{40}
      C_4(f) = \int_{\R^d}C_4(f;x)\tilde dx
      \eeq
      where
      \beq\label{C4}
      C_4(f;x) = C_{4,1}(f;x) + C_{4,2}(f;x) + C_{4,3}(f;x)
      \eeq 
            and 
      \bea
      C_{4,1}(f;x) &=& 2\oint_\Gamma(P-z)\sum_{\vert\beta\vert =4}\left(\partial^\beta_x(P-z)^2I_\beta^{(1)}\right)f(z)dz,\\
      C_{4,2}(f;x) &=& 2\sum_{\vert\alpha\vert=2}\Gamma(\alpha,0)\oint_\Gamma\left(\int_{\R^d}\partial^\alpha_\xi(\vert\xi\vert^2)
      \partial_x^\alpha\left(\frac{P-z}{L(z)}\right)K_2d\xi\right)f(z)dz,\\
      C_{4,3}(f;x) &=& 2\oint_\Gamma(P-z)\partial_x^\beta L(z)I_{\beta}^{(3)}f(z)dz.
          \eea
          Now we compute each term. After elementary but tedious computations we get the following results, using the notations:
          \bea
             \partial_j  &= & \frac{\partial}{\partial_{x_j}},\;\;\partial_j^2= \frac{\partial^2}{\partial^2_{x_j}},\;\; 
             \partial^2_{j,k} = \frac{\partial^2}{\partial^2_{x_j,x_k}} \\
              a_1 &=& \frac{1}{6}b_{6,2} -\frac{1}{3}b_{5,1}+\frac{1}{96}b_4\\
               a_2 &=& b_{6,1,1}
            \eea
          For $\fbox{d=5}$, we have 
          \bea
           C_{4,1}(f;x) &=& -20f(P)\Bigl[a_1\sum_{1\leq j\leq 5}\partial_j^4P +a_2\sum_{j<k}\partial_j^2\partial_k^2P\Bigr], \\
            &  &+  8f^\prime(P)\Bigl[a_1\sum_{1\leq j\leq 5}(\partial_j^2P)^2 + a_2\sum_{j<k}(\partial_{j,k}^2P)^2\Bigr],\non \\
             C_{4,2}(f;x) &=&\frac{E_1}{2}f^\prime(P) -\frac{E_2}{4}f^{\prime\prime}(P) +\frac{E_3}{12}f^{(3)}(P),\\
             C_{4,3}(f;x) &=&\frac{G_1}{2}f^\prime(P) -\frac{G_2}{4}f^{\prime\prime}(P) +\frac{G_3}{12}f^{(3)}(P).
             \eea
            where 
            \bea
            E_1 &=& (\triangle P)^2\Bigl(b_4-2b_5-2b_{5,1}+4b_{6,1}\Bigr),\\
            E_2 &=& \vert\nabla P\vert^2\triangle P\Bigl(b_4 +12b_6 -10b_5 -2b_{5,1}+16b_{6,1}-16b_{7,2}\Bigr),\\
            E_3 &=& \vert\nabla P\vert^4\Bigl(20b_6 -6b_5 -16b_7 +16b_{7,1} -12b_{6,1}\Bigr),\\
              G_1 &=& (\triangle P)^2\bigl(12b_{6,1} -2b_5\bigr) -16b_{7,2}\sum_j(\partial_j^2P)^2\\
              & & -32b_{7,1,1}\sum_{j\neq k}(\partial_j^2P)(\partial_k^2P)  -16b_{7,1,1}\sum_{j\neq k}(\partial_{j,k}^2P)^2,
            \non\\
            G_2 &=& \triangle P\vert\nabla P\vert^2\Bigl(24b_{6,1} -16b_{7,1} -4b_5 +4b_6\Bigr) 
             -32b_{7,2}\sum_{j}(\partial_jP^2)(\partial_jP)^2 \non \\ 
             & & -32b_{7,1,1}\sum_{j\neq k}(\partial_j^2P)(\partial_kP)^2
             -64b_{7,1,1}\sum_{j\neq k}(\partial_{j,k}^2)(\partial_jP)(\partial_kP), \\
                    G_3 &=& \vert\nabla P\vert^4\Bigl(12b_{6,1} -16b_{7,1} -2b_5 +4b_6\Bigr) 
               -16b_{7,2}\sum_{j}(\partial_jP)^4  \non\\
               & &  -48b_{7,1,1}\sum_{j\neq k}(\partial_jP)^2(\partial_kP)^2.
                         \eea
                   Finally, we get  
                                 \beq
           C_4(f;x) = A_0(x)f(P(x)) + A_1(x)f^\prime(P(x)) + A_2(x)f^{\prime\prime}(P(x)) + A_3(x)f^{(3)}(P(x)),
          \eeq
         \bea
         A_0(x) &=& -20a_1\sum_{j}\partial_j^4P -20a_2\sum_{j<k}\partial_j^2\partial_k^2P,\\
         A_1(x) &=& 8(a_1-b_{7,2})\sum_{j}(\partial_j^2P)^2 + 4(a_2-2b_{7,1})\sum_{j\neq k}(\partial_{j,k}^2P)^2 \\
         &  & + \frac{(\triangle P)^2}{2}\Bigl(b_4 -4b_5 -2b_{5,1} +16b_{6,1}\Bigr) 
            -16b_{7,1,1}\sum_{j\neq k}(\partial_j^2P)(\partial_k^2P), \non \\
         A_2(x) &=& \frac{\vert\nabla P\vert^2\triangle P}{4}\Bigr(14b_5 -16b_6 -b_4 +2b_{5,1} -40b_{6,1} +32b_{7,1}\Bigl)\\
         & &  +8b_{7,2}\sum_{j}(\partial_j^2P)(\partial_jP)^2   + 8b_{7,1,1}\sum_{j\neq k}(\partial_j^2P)(\partial_kP)^2 \non\\
          & &  +16b_{7,1,1}\sum_{j\neq k}(\partial_{j,k}^2P)(\partial_jP)(\partial_kP),\non\\
          A_3(x) &=& \frac{\vert\nabla P\vert^4}{12}\Bigl(24b_6 -8b_5 -16b_7\Bigr) -\frac{4}{3}b_{7,2}\sum_{j}(\partial_j P)^4 \non\\
          & & -4b_{7,1,1}\sum_{j\neq k}(\partial_jP)^2(\partial_kP)^2.
         \eea
         Using explicit computations (see Appendix), we get
          \bea
         A_0(x) &=& \frac{\pi^3}{24}\sum_{j}\partial_j^4P  -\frac{\pi^3}{48}\sum_{j<k}\partial_j^2\partial_k^2P,\\
         A_1(x) &=& -\frac{11\pi^3}{480}\sum_{j}(\partial_j^2P)^2 + \frac{\pi^2}{480}
         \sum_{j\neq k}(\partial_{j,k}^2P)^2 \\
         &  & -\frac{\pi^3}{160}(\triangle P)^2
            -\frac{\pi^2}{240}\sum_{j\neq k}(\partial_j^2P)(\partial_k^2P), \non \\
         A_2(x) &=& \frac{\pi^3}{96}  \nabla P\vert^2\triangle P +
         \frac{\pi^3}{160}\sum_{j}(\partial_j^2P)    (\partial_jP)^2  \\
         & &  +\frac{\pi^2}{480}\sum_{j\neq k}(\partial_j^2P)(\partial_kP)^2 
        +\frac{\pi^2}{240}\sum_{j\neq k}(\partial_{j,k}^2P)(\partial_jP)(\partial_kP),\non\\
          A_3(x) &=&-\frac{\pi^3}{576}\vert\nabla P\vert^4 -\frac{\pi^3}{960}\sum_{j}(\partial_j P)^4 
     -\frac{\pi^3}{960}\sum_{j\neq k}(\partial_jP)^2(\partial_kP)^2.
         \eea

          For $\fbox{d=7}$ we have 
          \beq
           C_4(f;x) = A_0(x)f(P(x)) + A_1f^\prime(P(x)), 
          \eeq
           \bea
          A_0(x) &=& \frac{\vert\nabla P\vert^2\triangle P}{2}
          \Bigl(14b_5 -16b_6 -b_4 +2b_{5,1} -40b_{6,1} +32b_{7,1}\Bigr)\\
         & &   +16b_{7,2}\sum_{j}(\partial_j^2P)(\partial_jP)^2 + 16b_{7,1,1}
         \sum_{j\neq k}(\partial_j^2P)(\partial_kP)^2 \non\\
          & & + 32b_{7,1,1}\sum_{j\neq k}(\partial_{j,k}^2P)(\partial_jP)(\partial_kP),\non\\
             A_1(x) &=& \frac{\vert\nabla P\vert^4}{2}\bigl(24b_6 -8b_5 -16b_7\bigr) -8b_{7,2}\sum_{j}(\partial_j P)^4 \non\\
           &  & -24b_{7,1,1}\sum_{j\neq k}(\partial_jP)^2(\partial_kP)^2.
\eea
 Using explicit computations as for $d=5$  (see Appendix), we get
          \bea
          A_0(x) &=&    \frac{\pi^4}{120}\sum_{j}(\partial_j^2P)(\partial_jP)^2 +
           \frac{\pi^3}{360}\sum_{j\neq k}(\partial_j^2P)(\partial_kP)^2 \non\\
           &  & 
           +\frac{\pi^3}{180}\sum_{j\neq k}(\partial_{j,k}^2P)(\partial_jP)(\partial_kP),\\
             A_1(x) &=& -\frac{\pi^4}{240}\vert\nabla P\vert^4 -\frac{\pi^4}{240}\sum_{j}(\partial_j P)^4 
            -\frac{\pi^3}{240}\sum_{j\neq k}(\partial_jP)^2(\partial_kP)^2.
\eea
  We should like to use these formulas  with     $f(\lambda) = (\lambda + t)^{-\mu}$,  $\mu>d(m+1)/m$ and $t>0$,
      to prove that $C^{(d)}_4(f) = \int_{\R^d}C_4(f;x)\tilde dx\neq 0$ ($d=5, 7$). \\
      With $f$ like above, we can see easily that $C^{(d)}_4(f)\neq 0$ for the following polynomials:\\
     $\boxed{1}$ $\di{P(x) =\sum_{1\leq j\leq d}\alpha_jx_j^m}$, $\alpha_j>0$, $1\leq j\leq d$, $d=5, 7$.\\
     $\boxed{2}$ $\di{P(x) =\sum_{1\leq j, k\leq d}\alpha_{j,k}x_jx_k}$, is a positive-definite quadratic form,
     $d=5, 7$\\
     $\boxed{3}$  $d=7$ and $P$ is convex.\\
     $\boxed{4}$  $d=5$,   $P$  is convex and satisfies the inequalities
     \bea
     \sum_{1\leq j<k\leq 5}\partial_j^2\partial_k^2 P  &\leq & 2\sum_{1\leq j\leq 5}\partial_j^4 P\\
     \sum_{j\neq k}\Bigl(\partial_{j,k}^2P\Bigr)^2  &\leq & 11\pi\sum_{1\leq j\leq 5}\Bigl(\partial_{j}^2P\Bigr)^2
     \eea
         For non-convex polynomials, we can check that $C^{(d)}_4(f)\neq 0$, $d=5, 7$,   for many examples with numerical computations, supporting  our conjecture that 
         for every elliptic polynomial $P$,  $C^{(d)}_4(f)\neq 0$, if  $d=5, 7$(see Appendix). 
    
      For $d=9, 11, $ it seems difficult to compute $C^{(d)}_6(f)$ by hand. We   need  more help from  symbolic  and numerical
       computations  to check our conjecture.

       \section{appendix}
       \subsection{Formulas for  $b_{j,k,\ell}$}
       We assume $d\geq 3$ and $2j-q>1$. 
 We have 
      \beq
        \int_0^{+\infty}\frac{r^q}{(1+r^2)^j}dr = \frac{1}{2}B(\frac{q+1}{2}, j-\frac{q+1}{2})
        \eeq
        where 
        $$
        B(x,y) = \int_0^1t^{x-1}(1-t)^{y-1}dt =\frac{\Gamma(x)\Gamma(y)}{\Gamma(x+y)}
        $$
        So  computing in polar coordinates,
        $$
        b_j(d) = \frac{\pi^{d/2}\Gamma(j-d/2)}{\Gamma(j)}
        $$
        Now, by elementary computations we get easily
        \bea
        b_{j,1}(d) &=& \frac{1}{d}(b_{j-1}(d) - b_j(d))\\
        b_{j,2}(d) &=& B(5/2, j-\frac{d+4}{2})b_j(d-1)\\
        b_{j,1,1}(d) &=& \frac{1}{8}B(3, j-\frac{d+4}{2})b_j(d-2).
        \eea
        
        \subsection{Numerical computations for $C_4(f)$}
        The following computations has been performed by Guy Moebs, Research Engineer, Laboratoire Jean-Leray, CNRS-University of Nantes.
        
         The method used to compute multi-dimensional integrals is Monte-Carlo, with a cut-off of the domain
        to reduce it  in a bounded  domain  fitting  with the behaviour of the polynomial $P$.\\
        In each example, 100 simulations are computed with at least $10^9$   events.
        
  \noi
   Example.1 $d = 5$, polyn\^ome $P(\bf{x}) = {\displaystyle \sum_{j=1}^d {x_j}^4} + \alpha {x_1}^2 {x_2}^2$
$$
\begin{array}{|r|r|}
\hline
\alpha & C_4(f)\\
\hline
7    &   1~428 \\
10   &   1~515\\
100  &   9~237\\
1000 & 235~115\\
\hline
\end{array}
$$

\noi
Example.2  $d = 7$, polyn\^ome $P(\bf{x}) = {\displaystyle \sum_{j=1}^d {x_j}^4} + \alpha {x_1}^2 {x_2}^2 + \beta {x_3}^2 {x_4}^2$ 
$$
\begin{array}{|r|r|r|}
\hline
\alpha & \beta & C_4(f)\\
\hline
7    &   7   &        409 \\
7    &  10   &        423 \\
7    &  100  &      1~806 \\
7    &  1000 &     39~646 \\
\hline
10   &   10  &        434\\
10   &   100 &      1~705\\
10   &   1000&     36~724\\
\hline
100  &  100 &       1~755\\
100  &  1000&      19~587\\
\hline
1000 & 1000 &      18~270\\
\hline
\end{array}
$$

\noi
Example.3  $d = 5$, polyn\^ome $P(\bf{x}) = {\displaystyle \sum_{j=1}^d {x_j}^6} + \alpha {x_1}^2 {x_2}^4 + \beta {x_3}^2 {x_4}^4$
$$
\begin{array}{|r|r|}
\hline
(\alpha, \beta) & C_4(f)\\
\hline
(100,10)    &   11~732 \\
\hline
\end{array}
$$

\end{document}